\documentclass
[twocolumn,superscriptaddress,showpacs,aps,prx,longbibliography]{revtex4-1}%
\usepackage{amsfonts}
\usepackage{amsmath}
\usepackage{amssymb}
\usepackage{bm}
\usepackage{graphicx}
\usepackage{txfonts}
\usepackage{tabularx}
\usepackage{multirow}
\usepackage{subfigure}
\usepackage{siunitx}
\usepackage{float}
\usepackage{mathrsfs}
\usepackage{braket}
\usepackage{xcolor} 
\usepackage[colorlinks,linkcolor=blue,anchorcolor=blue,citecolor=blue,urlcolor=blue,driverfallback=dvipdfm]%
{hyperref}%
\begin{document}

\preprint{APS/123-QED}

\title{Non-Abelian Gauge Field Optics in the Time Domain
}

\author{Yucheng Lai}
\affiliation{
Center for Quantum Matter, Zhejiang University, Hangzhou 310027,  China}
\affiliation{State Key Laboratory of Semiconductor Physics and Chip Technologies, Institute of Semiconductors, Chinese Academy of Sciences, Beijing 100083, China}
\affiliation{College of Materials Science and Opto-electronic Technology, University of Chinese Academy of Sciences, Beijing 100049, China}
\author{Yongliang Zhang}
\email{ylzhanglight@semi.ac.cn}
\affiliation{State Key Laboratory of Semiconductor Physics and Chip Technologies, Institute of Semiconductors, Chinese Academy of Sciences, Beijing 100083, China}
\affiliation{College of Materials Science and Opto-electronic Technology, University of Chinese Academy of Sciences, Beijing 100049, China}
\author{Kai Chang}
\email{kchang@zju.edu.cn}
\affiliation{
Center for Quantum Matter, Zhejiang University, Hangzhou 310027,  China}%

\begin{abstract}
Artificial gauge fields open up burgeoning opportunities for wave engineering in different disciplines. So far, previous works have mostly focused on synthesizing spatial gauge fields, where the pseudo-magnetic fields lie at the heart of these phenomena. In this Letter, we generalize the paradigm of gauge field optics to the time domain by using time-varying media with rotating anisotropy. Dual to its spatial counterpart, the temporal gauge field induces a pseudo-electric field for optical pulses, leading to the spin-dependent longitudinal shift and Zitterbewegung for both trajectory and frequency. In addition, we analyze the temporal non-Abelian interference effect induced by temporally bounded non-Abelian gauge field media, which results in the temporal spin-precession and the temporal analogy of the non-Abelian Aharonov-Bohm effect. Our work not only fills the gap between synthetic gauge fields and time-varying physical systems, but also provides a fundamentally new approach for manipulating light with time-varying media.

\end{abstract}

\maketitle


\textit{Introduction.}---Gauge field is one of the central concepts in unifying a variety of physical phenomena ranging from the fundamental forces of nature \cite{gf2} to the exotic manipulation of classical and quantum waves, including acoustics \cite{gfa1,gfa3}, electronics \cite{gfm1,gfm2,gfm3}, and cold atoms \cite{gfc1,gfc2}. According to the commutativity of gauge groups, gauge field theories can be classified into U(1) Abelian theory for electromagnetism \cite{ED} and non-Abelian gauge theories, which describe the SU(N) or Poincare invariance of elementary particles \cite{gf2}.
Recently, a growing interest is devoted to synthesizing U(1) \cite{gfo7,gfo6,gfo17,gfo3,gfo8,gfo18,gfo13} and SU(2) \cite{gfo4,gfo16,gfo11,gfo20} gauge fields for light with waveguides \cite{gfo7}, metamaterials \cite{gfo6,gfo17}, photonic crystals \cite{gfo3,gfo8,gfo18}, optical cavities \cite{gfo4,gfo16} and synthetic dimensions \cite{gfo13,gfo11,gfo20}. The emergent gauge structures provide a powerful paradigm for controlling the phase, propagation, and polarization of light in an unprecedented manner, enabling a variety of novel physical effects such as extraordinary waveguiding and trapping of light \cite{WD1,WD2}, optical spin-orbit interaction \cite{SOC1,SOC3}, and topological photonics \cite{TP1,TP2}.

In fundamental physics, the interaction between gauge fields and Fermions originates from the preservation of gauge invariance of matter fields under the local transformation of an irreducible representation of a certain Lie group \cite{gf2}. To create the associated non-integrable phase factor, specific hopping or material modulation in discrete or continuous systems is required \cite{gf3}. For instance, the simple U(1) gauge field has been created by introducing elastic strain in periodic lattices \cite{gfo14,gfo15,gfo19} or tilted material anisotropy in continuous media \cite{gfo6}. By replacing the isospin doublet with two orthogonal optical modes, more generic SU(2) non-Abelian gauge fields for light are synthesized in materials with in-plane duality symmetry \cite{gfo5} and later in biaxial dielectric materials with the real-space rotated permittivity tensor \cite{gfo23}, where the Zitterbewegung (ZB) and non-Abelian Aharonov-Bohm (AB) effects arise from the non-Abelian Lorentz force. Very recently, it was proposed to generate non-Hermitian topology with non-Abelian gauge fields, which enable non-Hermitian topological phase transition in one-dimensional (1D) systems lacking gauge flux and break the gauge invariance of Wilson loops in 2D \cite{nhgfo1}.

While significant effort has been made concerning the fundamentals and applications of artificial gauge fields, previous works mainly focused on synthesizing spatial gauge fields, which simulate matrix-valued gauge fields akin to usual electric or magnetic fields for wave systems \cite{gfo14,gfo15,SOC1,SOC3}. Recently, time-varying media enabled by the external temporal modulation have emerged as a new paradigm for unprecedented wave engineering, such as temporal scatterings \cite{TVM3,TVM12,TVM13}, photonic time crystals \cite{TVM14,TVM15,TVM16}, magnetic-free non-reciprocity \cite{TVM8,TVM9,TVM10}, and synthetic motions \cite{TVM6}. Except for few works synthesizing Abelian pseudo-magnetic fields with time-varying systems \cite{gfo8,ab2,gfo21,gfo22}, the role of time--as a fundamental physical degree of freedom dual to space--has not been fully addressed in the context of emergent gauge fields within a common ground with other physical quantities.

In this Letter, we present a time-space dual framework for the emergent non-Abelian gauge fields by extending the paradigm of non-Abelian gauge field optics to the time domain. Using spatially homogeneous time-varying optical media with rotating anisotropy, we demonstrate the synthesis of temporal Abelian and SU(2) gauge fields for light, which exhibit novel physical effects dual to their spatial counterparts. As a demonstration, we reveal that the temporal gauge field creates a pseudo-electric field for optical pulses, leading to spin-dependent longitudinal shift and ZB effect for both trajectory and frequency. Additionally, we show the temporal interference effect by the temporal spin-precession and the non-commutative temporal interference for light propagating through temporally bounded gauge field media. We further propose the temporal analog of the non-Abelian AB effect induced by sequentially temporal non-Abelian gauge field slabs. As evidenced by the examples, our work bridges the gap between gauge field optics and time-varying optical systems, and establishes a fundamentally new methodology towards temporal/frequency engineering of electromagnetic fields by using time-varying media.

\textit{Temporal SU(2) gauge fields.}---To synthesize non-Abelian gauge fields in the time domain, we consider the propagation of light in a spatially homogeneous anisotropic medium whose permittivity tensor depends on time. Without loss of generality, we assume that a plane wave with the initial frequency $\omega_0$ propagates along the $z$-axis in a general Hermitian medium described by 

\begin{equation}\label{1}    
\bm{\epsilon}(t)=\epsilon_r[\mathbb{I}_{2\times 2}+\bm{\Gamma}(t)],\enspace\bm{\Gamma}(t)=
\begin{bmatrix}
  a(t) & b(t)+ig(t)  \\
  b(t)-ig(t) & -a(t)  \\
\end{bmatrix}.
\end{equation}
Here, $\epsilon_r$ is the isotropic part of the permittivity tensor, $\mathbb{I}_{2\times 2}$ is the 2D identity matrix, $a(t)$ and $b(t)$ describe the anisotropy of the medium which rotates in time if $a(t)/b(t)\neq\mathrm{const}$, and the gyration vector $\bm{g}(t)=g(t)\bm{e}_z$ represents the time-varying magneto-optical effect with $\bm{e}_z$ denoting the unit vector along the $z$-axis. Experimentally, such time-varying anisotropic media can be realized by the redistribution of electrons induced by ultrafast laser at optical frequencies \cite{sm}.

For simplicity, we assume the anisotropic tensor $\bm{\Gamma}(t)$ is weak ($|\bm{\Gamma}(t)|\ll 1$) and changes slowly over time. By neglecting high-order terms proportional to $\frac{d\Gamma_{ij}}{dt}$ and $\frac{d^2\Gamma_{ij}}{dt^2}$, the wave equation of light can be written as (see details in the supplementary material \cite{sm})
\begin{equation}\label{2}    
\mathcal{H}\bm{\mathrm{E}}=
\bar{\nabla}^{2}\bm{\mathrm{E}}+\left (i\frac{\partial}{\partial \bar{t}}+\mathcal{A}_t\right )^2\bm{\mathrm{E}}=0
.
\end{equation}
Here, $\bar{\nabla} \equiv\lambda_0\nabla$ is the dimensionless gradient operator with $\lambda_0=c/(\omega_0\sqrt{\epsilon_r})$, $\bar{t} \equiv\omega_0 t$ is the dimensionless time, $\bm{\mathrm{E}}=(E_x+ iE_y,E_x- iE_y)/\sqrt{2}$ is the two-component electric field wave function. $\mathcal{H}$ in Eq. (\ref{2}) resembles the Hamiltonian of a non-relativistic particle moving in a temporal non-Abelian gauge field, where $\mathcal{A}_t\equiv \bm{\bm{\mathrm{\Omega}}\cdot \hat{\sigma}}$ is the scalar SU(2) gauge potential with $\bm{\mathrm{\Omega}}=(a,-b,g)/2$ and $\hat{\bm{\sigma}}=(\hat{\sigma}_1,\hat{\sigma}_2,\hat{\sigma}_3)$ denoting the Pauli matrices. Here, the matrix-valued $\mathcal{A}_t$ is non-commutative ($[\mathcal{A}_t(t_1),\mathcal{A}_t(t_2)]\neq 0$) due to the presence of Pauli matrices. Unlike previous works on the spatial gauge field $\bm{\mathcal{A}}_s$ which creates a pseudo-magnetic field by $\bm{\mathcal{B}}=\nabla\times\bm{\mathcal{A}}_s$ and modifies the mechanical momentum $\bm{p}=[\bm{e}_z+\bm{\mathcal{A}}_s]/\lambda_0$, the temporal gauge field $\mathcal{A}_t$ is a time-dependent scalar potential which modifies the frequency of light according to $\braket{\omega}=\omega_0[1+\braket{\mathcal{A}_t}]$ with $\langle G \rangle=\bra{\bm{\mathrm{E}}}G\ket{\bm{\mathrm{E}}}/\braket{\bm{\mathrm{E}}|\bm{\mathrm{E}}}$ denoting the expectation value of $G$.

\textit{Non-Abelian evolution of optical pulses.}---To investigate the physical effects of the temporal gauge field, we first study the motion of an optical pulse in the temporal non-Abelian gauge field media. For simplicity, we consider an optical pulse propagating along the $z$-axis $\bm{\mathrm{E}}=\bm{A}(\bar{z},\bar{t})e^{i\bar{z}-i\bar{t}}$. The propagation equation of the pulse's envelope $\bm{\mathrm{A}}$ is given by
\begin{equation}\label{3}    
-2i\frac{\partial \bm{A}}{\partial \bar{z}}=\mathcal{H}'\bm{A},\quad\mathcal{H}'= -\frac{\partial^2}{\partial \xi^2} +\mathcal{A}_t(\xi),
\end{equation}
where $\xi=\bar{t}-\bar{z}$ is the relative coordinate in the pulse's reference frame moving with the group velocity $c/\sqrt{\epsilon_r}$ in the medium without the gauge field. $\mathcal{H}'$ describes the evolution of an optical pulse in a potential $\mathcal{A}_t$. In the geometrical optics limit, the propagation of an optical pulse in a weakly anisotropic medium can be described by the motion of a classical particle with intrinsic spin and color degrees of freedom. Here, the color represents the frequency of light. The Heisenberg equations of motion of the pulse read,
\begin{eqnarray}  
\langle\dot{\xi}\rangle&=&-i\left\langle \left [\xi,\mathcal{H}'\right ] \right\rangle=-\langle\omega\rangle,\enspace\enspace \label{4}\\ 
\langle\dot{\omega}\rangle&=&-i\left\langle \left [i\frac{\partial}{\partial \xi},\mathcal{H}'\right ] \right\rangle=\langle\hat{\mathscr{E}}\rangle,\enspace  \label{5}\\
\dot{\bm{s}}&=&-i\left\langle \left [\hat{\bm{\sigma}},\mathcal{H}'\right ] \right\rangle=2\bm{\mathrm{\Omega}}\times\bm{s} \label{6},
\end{eqnarray}
where the over-dot denotes the derivative with respect to $\bar{z}$, $\bm{s}=\langle \hat{\bm{\sigma}} \rangle$, and 
\begin{equation}\label{7}    
\hat{\mathscr{E}}=\frac{\partial \mathcal{A}_t}{\partial \xi}=\frac{\partial \mathcal{A}_t}{\partial \bar{t}}.
\end{equation}

Eqs. (\ref{4}-\ref{6}) describe the orbital motion of light and the evolution of the inner degrees of freedom in the non-Abelian gauge field medium. Eqs. (\ref{4}-\ref{5}) correspond to the conventional canonical equations $\bm{\dot{r}}=\bm{p},\bm{\dot{p}}=\bm{\mathcal{E}}$ for the motion of a charged particle in an electric field $\bm{\mathcal{E}}$, except that $\bm{p}$ is replaced by $\omega$. However, Eq. (\ref{4}) shows that the relative velocity $\langle\dot{\xi}\rangle$ is proportional to the central frequency rather than the momentum $\bm{p}$. Moreover, different from the usual electric field  $\bm{\mathcal{E}}=-\nabla \mathcal{A}_0$ where $\mathcal{A}_0$ is the scalar part of the spatial gauge field \cite{gfo5}, $\mathcal{A}_t$ accelerates the central frequency through the pseudo-electric field $\hat{\mathscr{E}}$ in the moving frame $\xi=\bar{t}-\bar{z}$. In addition, the pseudo-electric field is invariant under the gauge transformation \cite{sm}. Eq. (\ref{6}) shows that the pseudo-spin precesses around the axis $\bm{\mathrm{\Omega}}$ with the frequency $2|\bm{\mathrm{\Omega}}|$. Here, the pseudo-spin describes the polarization of light, where the two orthogonal polarizations correspond to a pair of antipodal points on the Bloch sphere. Note that during precession, the pseudo-spin along the direction of $\bm{\mathrm{\Omega}}$ is conserved. 

\begin{figure}[h]
\centerline{\scalebox{0.40}{\includegraphics{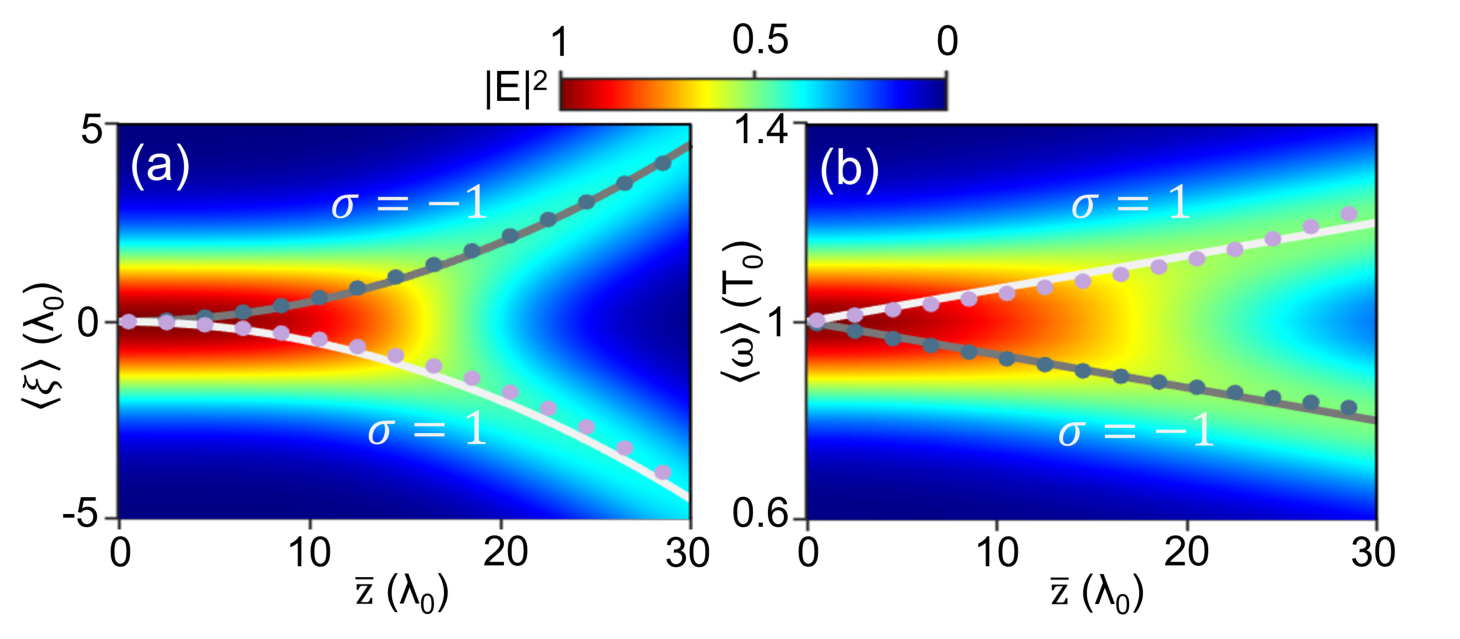}}}
\caption{\label{fig:1}
Motion of the optical pulse in a temporal Abelian gauge field $\mathcal{A}_t=0.01\hat{\sigma}_3\bar{t}$ with $\bm{\mathrm{E}}(\xi,\bar{z}=0)=\mathrm{exp}(-\xi^2/4)\bm{e}_x$. (a) Longitudinal shift of optical pulses. (b) Spin-dependent frequency shift of optical pulses. Here, $T_0=1/\omega_0$. The solid curves and density plots respectively represent the analytical results based on Eqs. (\ref{4}-\ref{6}) and the numerical calculation of Eq. (\ref{3}), while the dots are obtained by full-wave simulations.}
\end{figure}

\textit{Longitudinal shift and ZB in the time domain.}---We now investigate the motion of an optical pulse within an Abelian gauge field $\mathcal{A}_t=\bm{\mathrm{\Omega}}_a\cdot\hat{\bm{\sigma}}=M(\bar{t})\bm{n}_0\cdot\hat{\bm{\sigma}}$ with $M$ a real function of $\bar{t}$ and $\bm{n}_0$ a normalized vector. $\mathcal{A}_t$ is Abelian because the $\bm{n}_0\cdot\hat{\bm{\sigma}}$ components are commutative ($[\mathcal{A}_t(\bar{t}_1),\mathcal{A}_t(\bar{t}_2)]= 0$). For simplicity, we consider a linearly increased $\mathcal{A}_t=m\hat{\sigma}_3\bar{t}$ with $m$ a real number, which creates a pseudo-electric field
\begin{equation}\label{8}   
\langle\hat{\mathscr{E}}_a\rangle=\bm{\varepsilon}_a\cdot \bm{s},\quad \bm{\varepsilon}_a=[0,0,m].
\end{equation}
The electric field $\langle\hat{\mathscr{E}}_a\rangle$ induces a spin-dependent force for light. Specifically, spin-up ($\sigma=1$) and spin-down ($\sigma=-1$) photons experience opposite electric force $\langle\hat{\mathscr{E}}_\sigma\rangle=\sigma m$. This spin-dependent electric force arises from the coupling between the pseudo-spin and the trajectory of light in the time-varying anisotropic medium. Consequently, the trajectories of spin-up and spin-down photons split up upon propagation. In the electric field $\langle\hat{\mathscr{E}}_\sigma\rangle=\sigma m$, the longitudinal shift $\langle\xi_{\sigma}\rangle$ and the central frequency $\langle\omega_{\sigma}\rangle$ are given by
\begin{equation}\label{9}  
\begin{aligned} 
 \langle\xi_\sigma\rangle&=-\frac{1}{2}\sigma m \bar{z}^{2},\\
 \langle\omega_\sigma\rangle&=1+\sigma m\bar{z}.
 \end{aligned}
\end{equation}
In contrast to the transverse SHE \cite{SOC1,SOC3,SOC9}, the temporal gauge field leads to a spin-dependent longitudinal shift, which relates to the spin time delay $\braket{\Delta T_{\sigma}}=\langle\xi_\sigma\rangle/c$. Moreover, the temporal gauge field modifies the frequency of spin-up and spin-down photons oppositely, and therefore the central frequencies of the circularly-polarized light split upon propagation. As shown in Figs. \ref{fig:1}(a) and \ref{fig:1}(b), full-wave simulations of the pulse centroid and the frequency by using a finite element solver agree with the analytical results. For an arbitrary Abelian gauge field $\bm{\mathrm{\Omega}}_a=M(\xi)\bm{n}_0\cdot\hat{\bm{\sigma}}$, the longitudinal shift and frequency of two eigenstates of $\bm{n}_0\cdot\hat{\bm{\sigma}}$ are determined by $M$. Moreover, if a spatial gauge field is further introduced by a space-varying permittivity $\epsilon_r(\bm{r})$, the optical pulse exhibits asymmetrical SHE \cite{sm}.

\begin{figure}[h]
\centerline{\scalebox{0.4}{\includegraphics{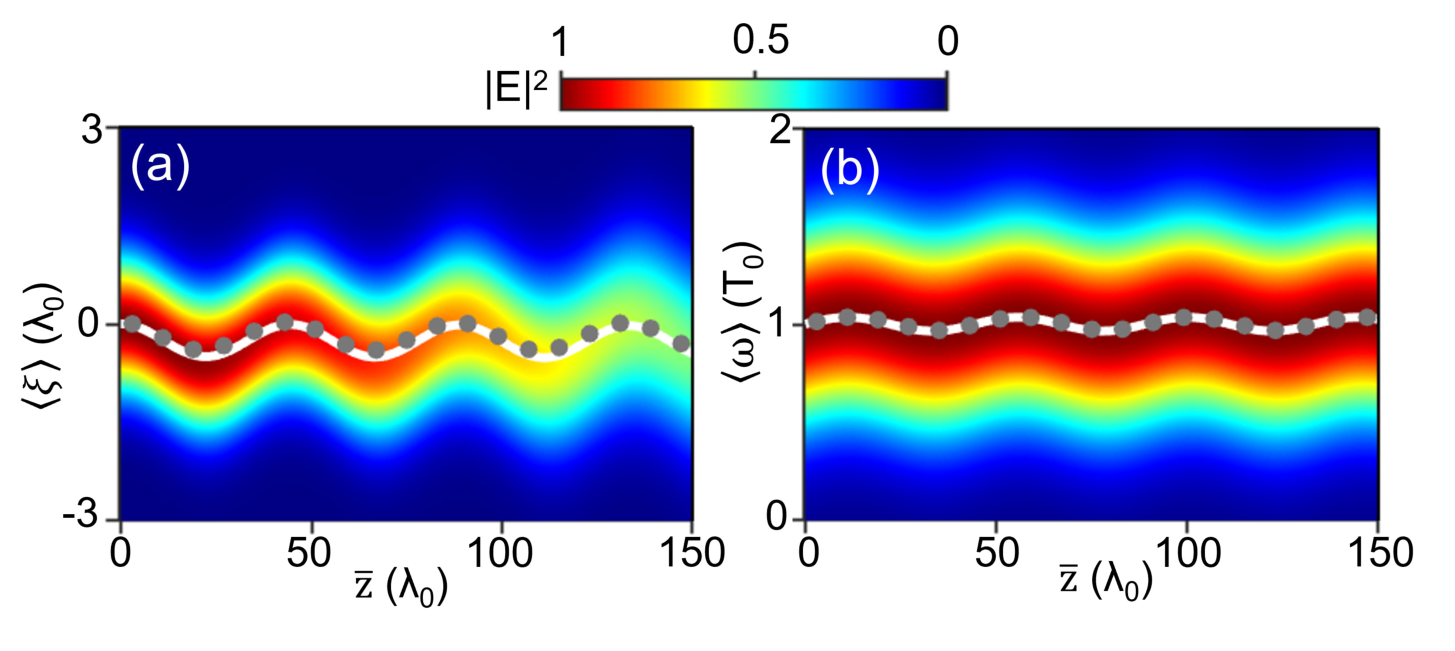}}}
\caption{ZB of the optical pulse in the temporal non-Abelian gauge field $\mathcal{A}_t=0.1\cos{0.1\xi}\hat{\sigma}_3+0.1\sin{0.1\xi}\hat{\sigma}_1$ with $\bm{\mathrm{E}}(\xi,\bar{z}=0)=\mathrm{exp}(-\xi^2)\bm{e}_+$. (a) Longitudinal ZB. (b) Frequency ZB. The solid curves and density plots respectively represent the analytical results based on Eqs. (\ref{4}-\ref{6}) and the numerical calculation of Eq. (\ref{3}), while the dots are obtained by full-wave simulations.}

\label{fig:2}
\end{figure}

We now consider a non-Abelian gauge field medium of the form $\mathcal{A}_t=\bm{\mathrm{\Omega}}_n\cdot\hat{\bm{\sigma}}$ with $\bm{\mathrm{\Omega}}_n=[m\cos{k\xi},0,m\sin{k\xi}]$, whose optical axis rotates in time with a constant magnitude $|\bm{\mathrm{\Omega}}_n|$. Compared with the Abelian case, it generates a non-Abelian pseudo-electric field
\begin{equation}\label{10}    
\langle\hat{\mathscr{E}}_n\rangle=\bm{\varepsilon}_n\cdot \bm{s},\quad \bm{\varepsilon}_n=[-mk\sin{k\xi},0,mk\cos{k\xi}].
\end{equation}
As in the Abelian case, the pseudo-spin $\bm{s}$ precesses around $\bm{\mathrm{\Omega}}_a$ with the frequency $2|\bm{\mathrm{\Omega}}_a|$. However, for the non-Abelian gauge field, the direction of $\bm{\mathrm{\Omega}}_n$ changes in time and the precession of $\bm{s}$ along the rotating $\bm{\mathrm{\Omega}}_n$ induces a periodic variation in $\langle\hat{\mathscr{E}}_n\rangle$. Based on Eqs. (\ref{4}-\ref{5}), this cyclic electric field induces a trembling trajectory for the pulse along the propagation direction. Unlike the transverse ZB in the spatial non-Abelian gauge field \cite{gfo5} and photonic lattices \cite{ZB1}, spin-up and spin-down photons exhibit opposite longitudinal ZB \cite{sm}. Moreover, $\langle\hat{\mathscr{E}}_n\rangle$ causes a periodic oscillation for the central frequency of the pulse, which is dual to the previously reported ZB of momentum \cite{gfo5}. The oscillation frequency of ZB is determined by the gauge field strength $|\bm{\mathrm{\Omega}}_n|=|m|$ and its periodicity $2\pi/k$. If $|k|\ll |m|$, the frequency approaches the precession frequency of the pseudo-spin $2|\bm{\mathrm{\Omega}}_n|$. Note that $2|m|$ is equal to the frequency difference of the orthogonal eigenstates $\ket{+}=[\mathrm{sin}(u),-\mathrm{cos}(u)]^{\mathrm{T}}$ and $\ket{-}=[\mathrm{cos}(u),\mathrm{sin}(u)]^{\mathrm{T}}$ with $u=(k\xi-\pi/2)/2$, revealing that ZB originates from the beating effect. From Fig. \ref{fig:2}, both the analytical trajectory and the central frequency are consistent with the full-wave simulation results. 

\textit{Temporal Abelian interference effect.}---As a second example, we consider the propagation of light in temporally bounded media, which have abrupt time interfaces but are homogeneous in space \cite{TVM4}. At a time interface, time reversal symmetry is broken while spatial symmetry is preserved, leading to the time reflection of part energy of a traveling wave associated with time reversal ($t\to-t$). As a result, the superposition of the forward and backward waves results in a temporal interference pattern $\mathrm{cos}(\bar{t})$, where the intensity oscillates periodically with the frequency $\omega_0$. Here, we consider a temporal slab where the spatially homogeneous gauge field medium occurs in the time interval from $0$ to $\Delta t$. At a fixed point, a plane wave acquires a phase factor in the gauge field slab
\begin{equation}\label{11}  
\begin{aligned} 
 &\qquad\bm{\mathrm{E}}(\Delta t)=e^{-i\Delta t}\mathcal{R}\bm{\mathrm{E}}(0).&\\
 \end{aligned}
\end{equation}
Here, $\mathcal{R}=\mathcal{T}\mathrm{exp}[i\int_0^{\Delta t}\mathcal{A}_t(\tau) d\tau]$ with $\mathcal{T}$ denoting the chronological operator, which differs from the phase factor a charged particle gain in the spatial gauge field \cite{ab4}.

\begin{figure}[h]
\centerline{\scalebox{0.39}{\includegraphics{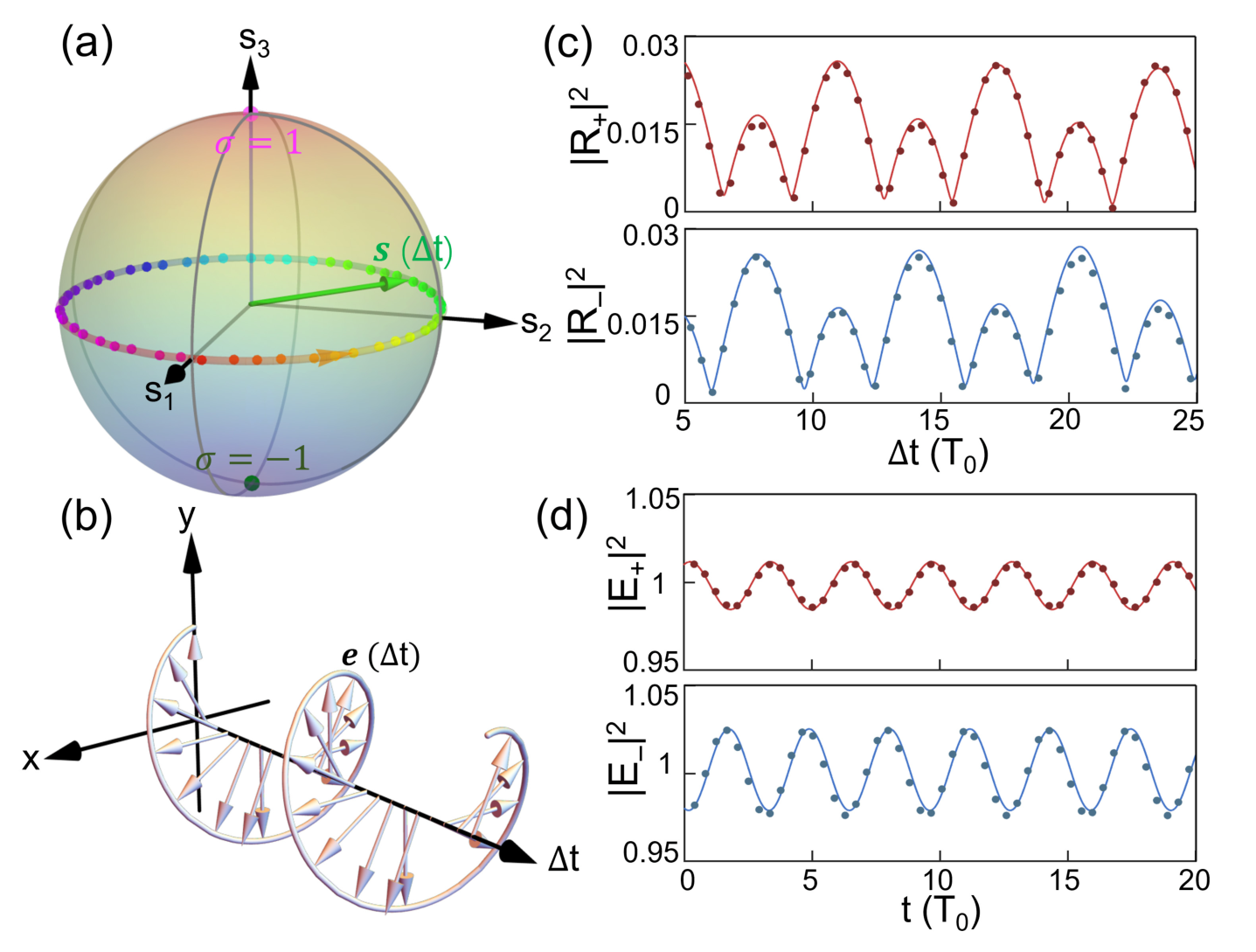}}}
\caption{ Temporal interference effect of the plane wave in a temporal Abelian gauge field slab $\mathcal{A}_t=0.01\hat{\sigma}_3$. (a) Temporal precession of the pseudo-spin on the Bloch sphere. (b) Evolution of the polarization vector. (c,d) The calculated results of (c) the reflectivity with different $\Delta t$ and (d) the intensity after scattering ($\bar{t}>\Delta t$). Here, the solid and dotted curves represent the analytical and the full-wave simulation results, respectively.}
\label{fig:3}
\end{figure}

As previously, we first analyze the evolution of polarization of a plane wave $\mathrm{exp}[i(\bar{z}-\bar{t})]$ scattered by a temporal Abelian slab during $\bar{t}\in [0,\Delta t]$:
\begin{equation*}  
\mathcal{A}_t=\bm{\mathrm{\Omega}}\cdot\hat{\bm{\sigma}},\quad\bm{\mathrm{\Omega}}= 
(0,0,m).
\end{equation*} 
Upon being scattered, light picks a spin-dependent phase factor $ \mathcal{\bm{R}}_a$:
\begin{equation}\label{12}    
 \mathcal{\bm{R}}_a(\Delta t)=
 \begin{bmatrix}
  e^{i\phi_B} & 0  \\
  0 & e^{-i\phi_B}  \\
\end{bmatrix}.
\end{equation}
Here, $\pm\phi_B$ with $\phi_B=m\Delta t$ are the opposite phases acquired by spin-up and spin-down photons. For an initially linearly polarized state $\bm{e}(0)$, the spin-up and spin-down components acquire opposite phases $\pm\phi_B$.  Thus, $\bm{e}(0)$ evolves into another linearly polarized state $\bm{e}(\Delta t)$. Fig. \ref{fig:3} (a-b) shows the analytical evolution of $\bm{s}(\Delta t)$ and the polarization vector $\bm{e}(\Delta t)$ with an initial polarization $\bm{e}_x$ based on Eq. (\ref{2}), which agree with the full wave simulation results. From Fig. \ref{fig:3}(a), the pseudo-spin $\bm{s}$ precesses around $\bm{\mathrm{\Omega}}$ with frequency $2|m|$. Moreover, the polarization vector $\bm{e}$ rotates counterclockwise along the propagation direction with $\Delta t$ [Fig. \ref{fig:3}(b)]. The rotation of $\bm{e}$ is the time-domain counterpart of the parallel transport of the polarization rotation upon propagation \cite{SOC3,SOC9}.

Compared with usual temporal slabs \cite{TVM4}, the amplitude of waves scattered by a gauge field slab is spin-dependent, which can be expressed as \cite{sm}:
\begin{equation}\label{13}    
E(\bar{t}>\Delta t)=T_\sigma e^{i(\bar{z}-\bar{t})}+R_\sigma e^{i(\bar{z}+\bar{t})}.
\end{equation}
Here, based on the first-order approximation of $|m|\ll 1$, the spin-dependent transmission and reflection coefficients are $T_\sigma=\cos{[(1+\sigma|m|)\Delta t]}+i(1+\alpha_\sigma^2)\sin{[(1+\sigma|m|)\Delta t]}/2\alpha_\sigma$ and $R_\sigma=-i(1-\alpha_\sigma^2)\sin{(\Delta t+\sigma |m|\Delta t)}/2\alpha_\sigma$, where $\alpha_\sigma=1/(1+\sigma|m|)$. The full-wave simulation results of the reflectivity and the temporal interference pattern for spin-up and spin-down photons are shown in Figs. \ref{fig:3}(c,d), which agree with the analytical results. Fig. \ref{fig:3}(c) demonstrates that the reflectivity of spin-up photons differs from that of spin-down ones, indicating the spin-related scattering in the time domain. Consequently, the temporal interference pattern from the supposition of reflected and transmitted waves is also spin-dependent [Fig. \ref{fig:3}(d)]. 

\begin{figure}[h]

\centerline{\scalebox{0.4}{\includegraphics{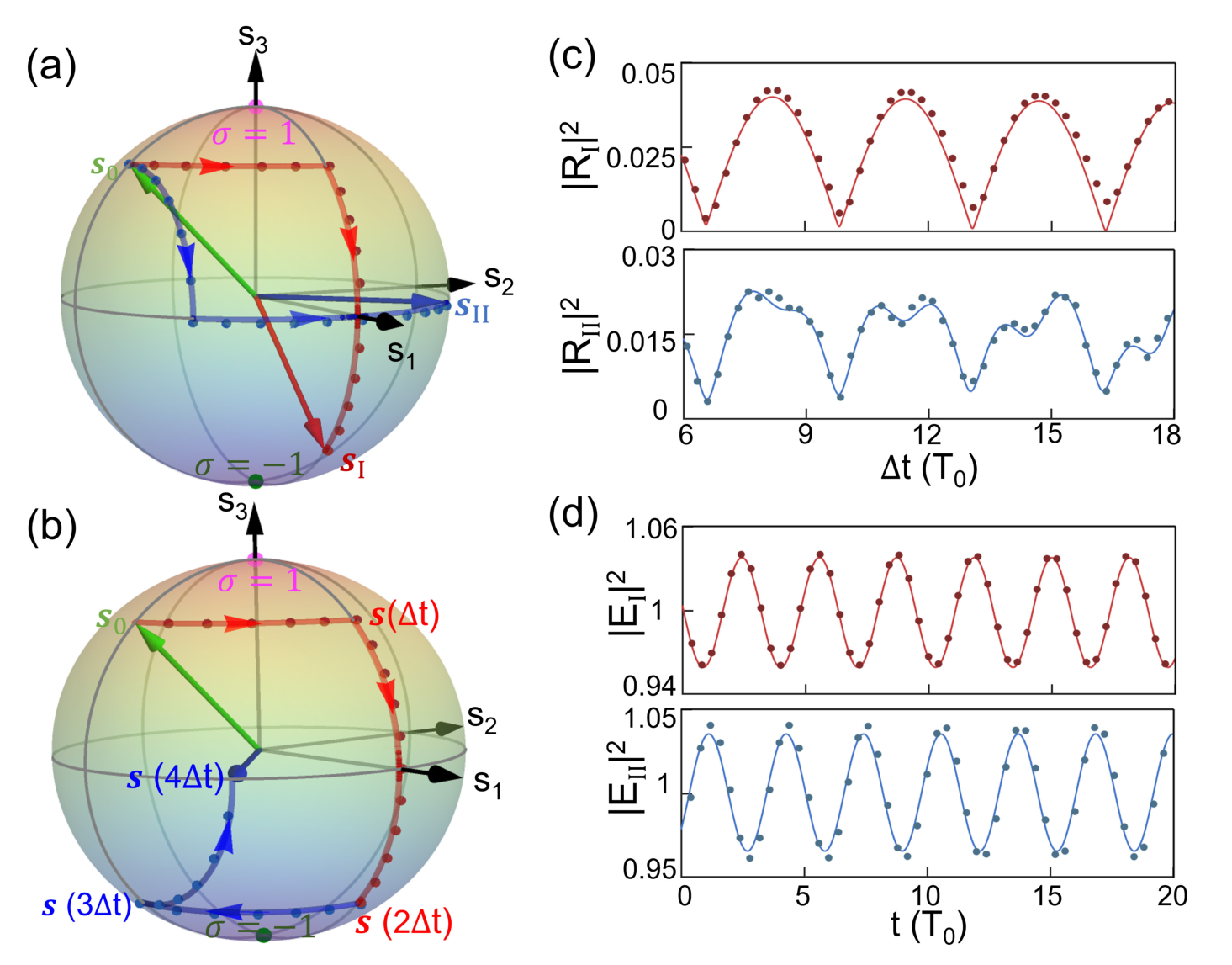}}}
\caption{Non-Abelian interference effect of the plane wave. (a-b) Temporal evolution of the pseudo-spin on the Bloch sphere in gauge fields (a) $\mathcal{A}_{\mathrm{I}}$ (red lines) and $\mathcal{A}_{\mathrm{II}}$ (blue lines), and (b) $\mathcal{A}_c$, with the initial spin $\bm{s}_0=(0,-1/\sqrt{2}, 1/\sqrt{2})$ and $\Delta t=\pi/2|m|$ with $m=0.02$. (c,d) The calculated results of (c) reflectivity and (d) intensity of plane waves after scattering ($\bar{t}>2\Delta t$). Here, the solid and dotted curves represent the analytical and the full-wave simulation results, respectively.}
\label{fig:4}
\end{figure}

\textit{Temporal non-Abelian interference effect.}---We now consider the temporal interference effect in a temporal non-Abelian gauge field. Due to the non-commutative nature, the temporal phase factor for the time sequence ($t_1\to t_2$) differs from the opposite time sequence ($t_2\to t_1$). To illustrate, we consider two non-Abelian temporal slabs $\mathcal{A}_{\mathrm{I}}=\bm{\mathrm{\Omega}}_{\mathrm{I}}\cdot \hat{\bm{\sigma}}$ and $\mathcal{A}_{\mathrm{II}}=\bm{\mathrm{\Omega}}_{\mathrm{II}}\cdot \hat{\bm{\sigma}}$ with
\begin{equation*}
\bm{\mathrm{\Omega}}_{\mathrm{I}}=\left\{
\begin{aligned} 
 &(0,0,m),\bar{t}\in [0,\Delta t]&\\
 &(0,m,0),\bar{t}\in [\Delta t,2\Delta t],&
 \end{aligned}
 \right.
\bm{\mathrm{\Omega}}_{\mathrm{II}}=\left\{
\begin{aligned} 
 &(0,m,0),\bar{t}\in [0,\Delta t]&\\
 &(0,0,m),\bar{t}\in [\Delta t,2\Delta t].&
 \end{aligned}
 \right.
\end{equation*}
where the two gauge fields $m\hat{\sigma}_3$ and $m\hat{\sigma}_2$ are arranged in reverse sequences. Upon being separately scattered by $\mathcal{A}_{\mathrm{I}}$ and $\mathcal{A}_{\mathrm{II}}$, a plane wave with the initial state $\bm{\mathrm{E}}(0)$ evolves into different states:
\begin{equation}\label{14}  
\begin{aligned} 
 &\enspace\bm{\mathrm{E}}_{\mathrm{I}}(\bar{t})=(T_{\mathrm{I}}e^{i(\bar{z}-\bar{t})}+R_{\mathrm{I}}e^{i(\bar{z}+\bar{t})})\mathcal{\bm{R}}_1\mathcal{\bm{R}}_2\bm{\mathrm{E}}(0),&\\
 &\bm{\mathrm{E}}_{\mathrm{II}}(\bar{t})=(T_{\mathrm{II}}e^{i(\bar{z}-\bar{t})}+R_{\mathrm{II}}e^{i(\bar{z}+\bar{t})})\mathcal{\bm{R}}_2\mathcal{\bm{R}}_1\bm{\mathrm{E}}(0).&
 \end {aligned}
 \end{equation}
Here, $\mathcal{\bm{R}}_1=\mathrm{exp}[i\Delta t m\hat{\sigma}_3] $ and $\mathcal{\bm{R}}_2=\mathrm{exp}[i\Delta t m\hat{\sigma}_2]$ are two temporal phase factors, $T_{\mathrm{I}}$ ($R_{\mathrm{I}}$) and $T_{\mathrm{II}}$ ($R_{\mathrm{II}}$) denote the transmission (reflection) coefficients of gauge field slabs $\mathcal{A}_{\mathrm{I}}$ and $\mathcal{A}_{\mathrm{II}}$. Eq. (\ref{14}) demonstrates the non-Abelian temporal interference for light in the non-Abelian gauge field. For instance, a right-handed elliptically polarized state $\bm{s}_0$ evolves into a left-handed elliptically polarized state $\bm{s}_{\mathrm{I}}$ after being scattered by $\mathcal{A}_{\mathrm{I}}$, while it evolves into a linearly polarized state $\bm{s}_{\mathrm{II}}$ after being scattered by $\mathcal{A}_{\mathrm{II}}$ [Fig. \ref{fig:4}(a)]. Besides, $\mathcal{\bm{R}}_1\mathcal{\bm{R}}_2$ and $\mathcal{\bm{R}}_2\mathcal{\bm{R}}_1$ generate different phases in the final states ($\bm{\mathrm{E}}_{\mathrm{I}}$, $\bm{\mathrm{E}}_{\mathrm{II}}$). The non-commutation of the non-Abelian gauge fields provides a tool to manipulate the polarization and phase of light.

AB effect is one of the prominent physical effects of the gauge field, which is associated with the non-integrable phase factor for particles traveling along a spatially closed path \cite{ab4}. Based on the non-Abelian slabs mentioned above, we propose the temporal analogy of the non-Abelian AB effect. Because time flows in only one direction, we consider the propagation of light through four temporal slabs consisting of consecutive gauge field media of  $m\hat{\sigma}_3\to m\hat{\sigma}_2\to -m\hat{\sigma}_3 \to -m\hat{\sigma}_2$, where each slab occurs in the same time interval $\Delta t$. The latter two slabs occur with the revered consequence of the former two, therefore forming a temporal closed loop. Upon propagation from $t=0$ to $4\Delta t$, a plane wave acquires a non-Abelian phase factor
\begin{equation}\label{15}  
\mathcal{U}_c=\mathcal{R}_2^{-1}\mathcal{R}_1^{-1}\mathcal{R}_2\mathcal{R}_1,
\end{equation}
$\mathcal{U}_c$ is the temporal counterpart of the usual spatial non-Abelian AB phase. In a temporal Abelian gauge field, the resulting phase factor satisfies $\mathcal{R}_a^{-1}\mathcal{R}_a=\mathbb{I}_{2\times2}$, and therefore the polarization state of the photon remains unchanged $\bm{s} (4\Delta t)=\bm{s} (0)$ after encircling a temporal close loop. However, the polarization state changes after being scattered by a non-Abelian close loop due to the non-commutation of $m\hat{\sigma}_3$ and $m\hat{\sigma}_2$. For example, a right-handed elliptically polarized state $\bm{s} (0)$ evolves into a linearly polarized state $\bm{s} (4\Delta t)$ after being scattered by $\mathcal{A}_c$ [Fig. \ref{fig:4}(b)]. The difference between $\bm{s} (0)$ and $\bm{s} (4\Delta t)$ demonstrates the non-Abelian AB effect in the time domain. This effect allows us to control the polarization of light at any point on the Bloch sphere, which is useful for novel compact polarization rotation devices.

Due to the non-Abelian nature, a plane wave respectively scattered by the reverse-sequential gauge field slabs $\mathcal{A}_{\mathrm{I}}$ and $\mathcal{A}_{\mathrm{II}}$ exhibits different time-varying amplitudes. Full-wave simulations of the reflectivity and scattering of plane waves by $\mathcal{A}_{\mathrm{I}}$ and $\mathcal{A}_{\mathrm{II}}$ shown in Figs. \ref{fig:4}(c,d) demonstrate that the reflectivity of the electromagnetic wave is distinct after being scattered by $\mathcal{A}_{\mathrm{I}}$ and $\mathcal{A}_{\mathrm{II}}$ for different $\Delta t$, which is consistent with the analytical results. Unlike the scattering by Abelian gauge field slabs, the reflectivity does not change periodically with $\Delta t$. Consequently, the temporal scattering patterns resulting from the superposition of forward and backward waves differ after being scattered by $\mathcal{A}_{\mathrm{I}}$ and $\mathcal{A}_{\mathrm{II}}$. Specifically, the time-varying intensity of the plane wave scattered by $\mathcal{A}_{\mathrm{I}}$  exceeds that of $\mathcal{A}_{\mathrm{II}}$ for $\Delta t=20$ $T_0$ [Fig. \ref{fig:4}(d)]. Moreover, the interference fringe of the plane wave scattered by $\mathcal{A}_{\mathrm{I}}$ shifts from that of $\mathcal{A}_{\mathrm{II}}$. This shift results from the phase difference between $\bm{\mathrm{E}}_{\mathrm{I}}$ and $\bm{\mathrm{E}}_{\mathrm{II}}$. Consequently, the non-Abelian temporal interference allows the full control of the phase, amplitude, and polarization of light. 

 \textit{Conclusions.}---In this Letter, we have extended the framework of gauge field optics to the time domain by using time-varying anisotropic media. Temporal Abelian and non-Abelian gauge fields are synthesized by anisotropic media with temporal inhomogeneity. Dual to their spatial counterparts, the temporal gauge fields act on optical pulses through the pseudo-electric field, leading to the spin-dependent longitudinal shift and ZB in the time domain. Moreover, it is demonstrated that the temporal non-Abelian gauge field induces the temporal interference effect for light scattered by several non-Abelian time interfaces. The extension to the time domain not only complements the framework of the artificial gauge field theory, but also provides opportunities to manipulate light in an unprecedented manner. In addition to controlling the trajectory and polarization, the temporal gauge field could also be used to explore a wide variety of non-Abelian physics, including topological physics, spin-orbit coupling and non-Abelian quantum computation with light. Our work can be extended to other wave systems, such as circuits, elastic, acoustic and spin waves.

This work is supported by the National Natural Science Foundation of China (NSFC; Grants No. 12488101, No. 12574058, and No. 92265203), the Strategic Priority Research Program of the Chinese Academy of Sciences (Grants No. XDB0460000 and No. XDB28000000), and the Quantum Science and Technology-National Science and Technology Major Project (Grants No. 2024ZD0300104 and No. 2021ZD0302600).

\end{document}